\documentclass[structabstract]{aa}
\usepackage{float}
\usepackage{lineno}
\usepackage{natbib}
\bibpunct{(}{)}{;}{a}{}{,}
\usepackage{subfigure}
\usepackage{graphicx}
\usepackage{epsfig}
\usepackage{amsmath}
\usepackage[varg]{txfonts}
\usepackage{lscape}
\usepackage{longtable}
\usepackage{txfonts}
\usepackage{color}

\makeatletter

\newcommand{\Rmnum}[1]{\expandafter\@slowromancap\romannumeral #1@}
\makeatother
\begin{document}
 %   \title{First global coronal free magnetic energy estimate}
   \title{ First use of synoptic vector magnetograms for global nonlinear force free coronal magnetic field models}
   \author{Tilaye Tadesse\inst{1}, T. Wiegelmann\inst{2},S. Gosain\inst{3}, P. MacNeice\inst{1},  \and Alexei A. Pevtsov\inst{3}
          }
\institute{Space Weather Laboratory, NASA Goddard Space Flight Center, Greenbelt, MD, USA\\
     \email{tilaye.tadesse.asfaw@nasa.gov; peter.j.macneice@nasa.gov}       
  \and Max-Planck-Institut f\"ur Sonnensystemforschung, Max-Planck-Strasse 2, 37191 Katlenburg-Lindau, Germany \\
   \email{wiegelmann@mps.mpg.de} 
 \and National Solar Observatory, Sunspot, NM 88349, USA.\\
      \email{sgosain@nso.edu; apevtsov@nso.edu}  
      }
  \date{Received ********** / Accepted ********}
%  \linenumbers
  \abstract
% context heading (optional)
{The magnetic field permeating the solar atmosphere is generally thought to provide the energy for much of the activity
seen in the solar corona, such as flares, coronal mass ejections (CMEs), etc. To overcome the unavailability of coronal 
magnetic field measurements, photospheric magnetic field vector data can be used to reconstruct the coronal field. Currently 
there are several modelling techniques being used to calculate three-dimension of the field lines into the solar atmosphere.
}
% aims heading (mandatory)
{For the first time, synoptic maps of photospheric vector magnetic field synthesized from Vector Spectromagnetograph (VSM) on 
Synoptic Optical Long-term Investigations of the Sun (SOLIS) are used to model the coronal magnetic field and estimate free 
magnetic energy in the global scale. The free energy (i.e., the energy in excess of the potential field energy) is one of the 
main indicators used in space weather forecasts to predict the eruptivity of active regions.
}
% methods heading (mandatory)
{ We solve the nonlinear force-free field equations using optimization principle in spherical geometry. The resulting 
three-dimensional magnetic fields are used to estimate the magnetic free energy content $E_\mathrm{free}=E_\mathrm{nlfff}-E_\mathrm{pot}$, 
i.e., the difference of the magnetic energies between the nonpotential field and the potential field in the global solar corona. 
For comparison, we overlay the extrapolated magnetic field lines with the extreme ultraviolet (EUV) observations by the Atmospheric Imaging 
Assembly on board SDO. 
}
% results heading (mandatory)
{For a single Carrington rotation 2121, we find that the global NLFFF magnetic energy density is 
$10.3\%$ higher than the potential one. Most of this free energy is located in active regions.
}

\vspace{-1.cm}

\titlerunning{Global nonlinear force free coronal magnetic field models}
\authorrunning {T. Tadesse \textit{et al}.}
   \maketitle
   
%================================================================================================================================
\section{Introduction}
%================================================================================================================================
In the solar coronal plasma, magnetic energy is the prime energy reservoir that fuels the dynamical evolution of 
eruptive events, but it remains an open question how the magnetic energy is released. The amount of energy 
associated with the magnetic field is much larger than other energy sources, and the dynamics of the coronal configuration 
is determined by the evolution of its magnetic field \citep{Forbes:2000,Low:2001}. 

Free energy is defined as excess of energy as compared with potential field. In fact, one can in principle release more energy 
than free energy by, for example, annihilating magnetic field. Also, numerically, one can get pre-post flare difference in energy 
larger than free magnetic energy if magnetic field at the photosphere changes (e.g., some flux elements disappear). It quantifies the 
energy deviation of the coronal magnetic field from its potential state \citep{Metcalf:2005,Regnier:2007,Aschwanden:2012}. The 
magnetic free energy is stored in the form of electric currents flowing along the magnetic field. Free magnetic energy of solar solar 
magnetic fields can be affected by several processes such as, e.g., photospheric shearing flows, magnetic flux emergence and magnetic 
reconnection \citep[e.g.,][]{Welsch2006,Fang:2012}. 

To understand the role that the magnetic field plays in energizing the solar corona, it is important to calculate 
the amount of free energy in order to quantify the energy budget in a catastrophic energy release event, as well as for 
estimating upper limits in forecasting individual events in real-time. Using various extrapolation techniques for the coronal
magnetic field under the assumption of force-free fields, the spatial and temporal evolution of the coronal magnetic free
energy during solar flares has been extensively studied \citep{Regnier:2006,Guo:2008,Jing:2010,Tilaye:2012,Meyer:2013}. From the 3D 
coronal magnetic configurations, we can derive the magnetic energy in the corona as:
\begin{equation}
   E_\mathrm{M}=\frac{1}{8\pi}\int_{V} \textbf{B}\cdot\textbf{B}\,\,r^{2}\sin\theta dr d\theta d\phi  \label{one}
\end{equation}
The free magnetic energy in spherical geometry is calculated by computing the nonpotential field $B_\mathrm{nlfff}(r,\theta,\phi)$ 
with a numerical nonlinear force-free field (NLFFF) code and a potential field $B_\mathrm{pot}(r,\theta,\phi)$ for the same 
photospheric boundary data, so that the difference of the magnetic field energy density in the coronal volume V
encompassing the active regions of interest can be quantified as $E_\mathrm{M_{free}}=E_\mathrm{M_{nlfff}}-E_\mathrm{M_{pot}}$.

Since the corona is optically thin, direct measurements of the 3-D magnetic field are very difficult to implement and interpret. 
Therefore, the present observations for the magnetic fields based on the spectropolarimetric method (the Zeeman and the 
Hanle effects) are limited to low layers of solar atmosphere (photosphere and chromosphere). Even if direct measurement 
techniques for the 3-D magnetic fields in the chromosphere and the corona have considerably improved in recent decades
\citep{Lin:2000,Lin:2004,Liu:2008}, further developments are needed before accurate data are routinely available. Thus, the problem 
of measuring the coronal field and its embedded electrical currents leads us to use numerical modelling to infer the field 
strength in the higher layers of the solar atmosphere from the measured photospheric field. 

Force-free extrapolation of photospheric magnetic fields is currently used as the primary tool for the modeling of coronal magnetic
fields. In this model assumption, the corona magnetic forces are dominant so that all non-magnetic forces like plasma pressure gradient 
and gravity can be neglected in the lowest order \citep{Gary:2001}. This implies that, if appreciable currents are present, these 
must be aligned with the magnetic field, since otherwise the resulting Lorentz forces could not be balanced by the nonmagnetic forces. 
The equilibrium structure of the static coronal magnetic field can be described using the force-free assumption as:
\begin{equation}
   (\nabla \times\textbf{B})=\alpha\textbf{B}\Rightarrow(\nabla \times\textbf{B})\times\textbf{B}=0 \label{two}
\end{equation}
\begin{equation}
    \nabla \cdot\textbf{B}=0 \label{three}
 \end{equation}
where $\textbf{B}$ is the magnetic field. The force-free parameter $\alpha$ of Equation~(\ref{two}) can be a function of position, 
but the combination of Equation~(\ref{two}) and (\ref{three}) ($\textbf{B}\cdot\nabla\alpha=0$) requires that $\alpha$ be 
constant along a given field line. Potential ($\alpha=0$) and linear force-free fields (whenever $\alpha$ is constant everywhere 
in the volume under consideration) can be used as a first step to model the general structure of magnetic fields in the solar corona. 
Practically the pre-eruptive magnetic fields are nonlinear force-free fields ($\alpha$ being a function of position) as supported by 
both observational and theoretical reasons. For details of those models we direct the readers to \citet{Wiegelmann:2012W}.

Nonlinear force-free field codes have been routinely applied to the reconstruction of the coronal field of a single 
active region using the Cartesian geometry. In that case, the curvature of the solar surface does not play an important role. Solar Dynamics 
Observatory (SDO) mission has made repeated observations of large scale events in which connections between widely separated 
active regions may play fundamental role \citep{Martens:2012}. Even before SDO, it was known that large-scale connectivity 
is important for solar eruptive and non-eruptive activity (e.g., studies of sympathetic flares, transequatorial loops \citep{Pevtsov:2000}, 
effects of distant active regions on large-scale coronal brightness \citep{Pevtsov:2001}, eruption of filaments triggered by 
remote flux emergence and evolution \citep[e.g.,][]{Balasubramaniam0:2011}. Therefore, this needs motivate us to implement a NLFFF 
procedure in spherical geometry \citep{Wiegelmann07,tilaye09,Tilaye:2011,Tilaye:2012,Tilaye:2012a,Guo:2012,Tadesse:2013a,Amari:2013}

In this study, we estimate the free magnetic energy for global corona using data from SOLIS/VSM. We compare 
the extrapolated potential and NLFFF magnetic loops with extreme ultraviolet (EUV) observations by the Atmospheric 
Imaging Assembly (AIA) on board SDO. This comparison helps to identify whether the NLFFF model reconstructs
the magnetic configuration better than the potential field model in the global scale. In this paper, we present 
some descriptions of the dataset used for analysis in Section 2. The spherical optimization procedure used for 
modeling 3-D magnetic field in global corona is presented in Section 3. Then, we present results of our studies 
in Section 4. A summary and discussions are finally presented in Section 5.
%================================================================================================================================
\section{Instrumentation and data set}
%================================================================================================================================
In this paper, we use the first ever synoptic vector magnetograms, i.e., Carrington maps of the three components of the magnetic 
field vector, the radial $B_r$, the poloidal $B_\theta$, and the toroidal $B_{\phi}$, to perform first global nonlinear force 
free field extrapolation based on optimization algorithm \citep{Wiegelmann07,tilaye09}. The synoptic vector field maps are derived 
using daily full disk photospheric vector magnetograms from Vector Spectromagnetograph (VSM) instrument of Synoptic Optical 
Long-term Investigations of the Sun (SOLIS), a synoptic observing facility \citep{Keller:2003,Balasubramaniam:2011}. 
Further detail about these synoptic maps and their properties can be found in \cite{Gosain:2013}. Here we briefly describe 
the instrument characteristics and the full disk vector field observations. 

VSM routinely obtains full disk magnetograms in photospheric and chromospheric lines as a part of the synoptic program of National
Solar Observatory i.e., NSO Integrated Synoptic Program (NISP). In order to obtain full disk photospheric vector magnetograms 
SOLIS/VSM measures Stokes $(S=I,Q,U,V)$ profiles in photospheric Fe I $630.15-630.25$ nm line pair. A single full disk scan 
(2048 scan lines) takes only about 20 minutes, thanks to the long slit of the spectrograph which intersects the solar disk 
from one limb to another in one shot. The spatial sampling is 1 arc-sec per pixel with square pixels. The spectral sampling 
is 2.4pm per pixel. A single Stokes set $(I,Q,U,V)$ per slit position is obtained typically in about 0.6 second. 
The telescope itself is designed to be free of instrumental polarization by employing symmetric optical configuration and 
performing polarization modulation just after the prime focus, after the slit. Dual beam analysis of polarization using polarizing 
beam splitter avoids seeing induced cross talks in the signal. The polarimeter calibration is done routinely to calibrate the Stokes 
vector for cross-talks. The signal to noise ratio (SNR) in the continuum of the Stokes profiles is typically $>1000$. The magnetic field 
vector is inferred from the calibrated Stokes profiles by performing inversion in the framework of Milne-Eddington model for stellar 
atmosphere following Unno-Rachkovsky formalism \citep{Skumanich:1987}. 

Only pixels with polarization signal above the threshold of $0.1\%$ of continuum intensity, Ic, are inverted to obtain the magnetic 
(field strength, inclination angle, and azimuth angle) and thermodynamic (e.g., Doppler width, Doppler velocity, source function, 
temperature) parameters. The threshold of $0.1\% $ of Ic corresponds to typical noise level in the continuum. Using this threshold 
avoids fitting profiles buried in the noise. 

Further details about instrument and pipeline reduction steps can be found elsewhere 
\citep[e.g.,][]{Jones:2002,Henney:2006,Balasubramaniam:2011}. The noise in SOLIS magnetograms is estimated 
to be few Gauss in longitudinal and 70 G in the transverse field measurements \citep{Tadesse:2013a}. 
The 180 degree azimuth ambiguity is resolved using a different (faster) ambiguity resolution method developed 
by \cite{Rudenko:2011}.
%===============================================================================================================================================
\begin{figure*}
\begin{center}
\includegraphics[viewport=0 10 500 710,clip,height=23.5cm,width=18.5cm]{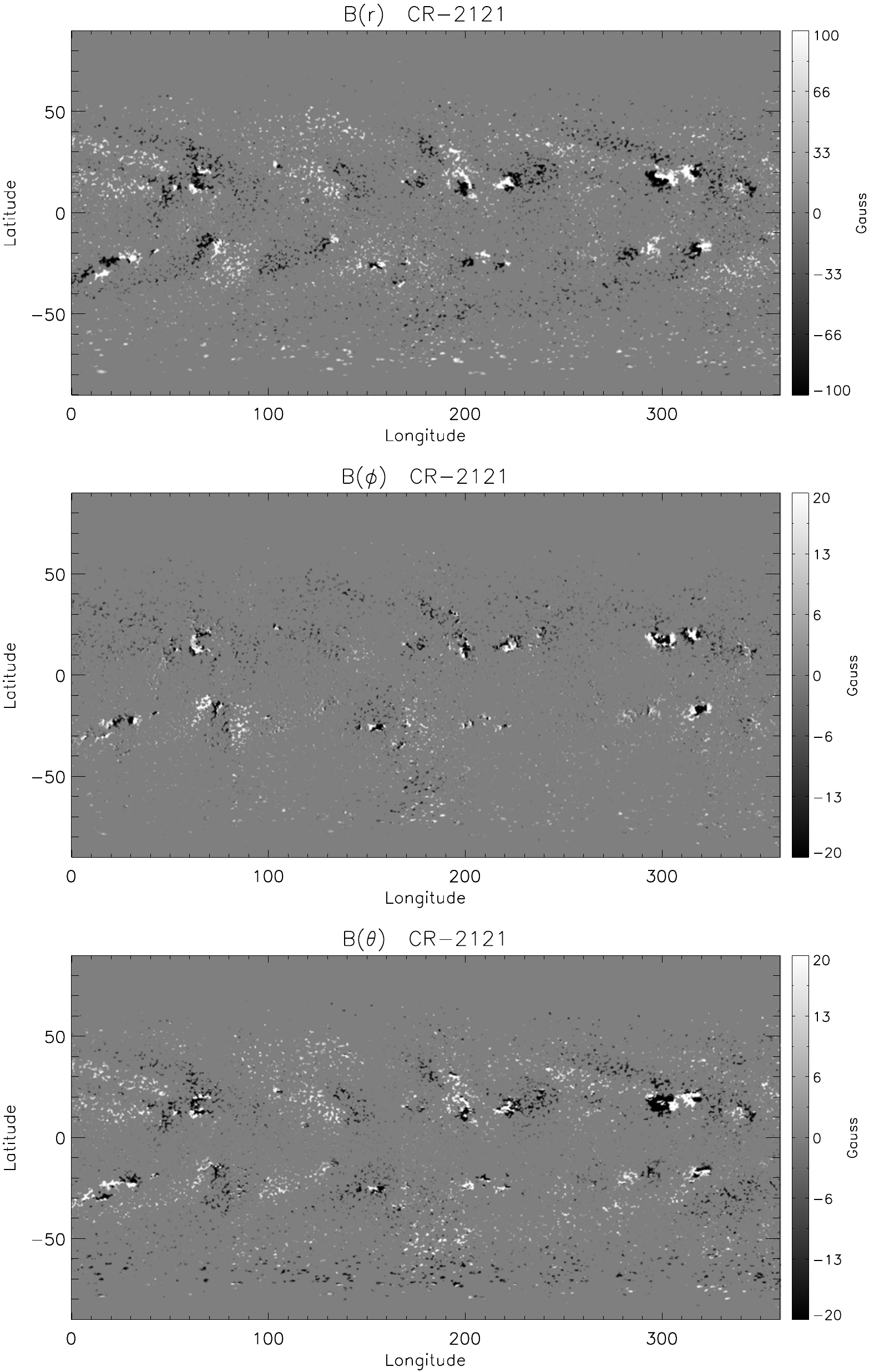}\\%{br_orig.eps}
\end{center}
\caption{Synoptic Carrington maps of the vector magnetic field components are shown for CR-2121. The panels from top to 
bottom show the distribution of the $B(r)$, $B(\phi)$ and $B(\theta)$ components, respectively. The $B_{r}$ map is scaled 
between $\pm$ 100 G, and the $B_{\phi}$ and $B_{\theta}$ maps are scaled to $\pm$ 20 G. The positive values of $B_{r}$, 
$B_{\phi}$ and $B_{\theta}$ point, respectively, upward, to the right (westward) and southward.}
\label{fig1}
\end{figure*}

%================================================================================================================================
\section{Magnetic field modeling}
%================================================================================================================================
For modelling the coronal magnetic field in a global scale, we use the variational principle originally proposed by \cite{Wheatland:2000}.
\cite{Wiegelmann07} has developed an optimization method to reconstruct the NLFFF for global solar corona by minimizing an 
objective functional $L$ that combines Lorentz forces and the divergence of the magnetic field in spherical geometry. The code has been tested 
with semi-analytic force-free solutions \citep{Low:1990}. If the functional is minimized to zero, Equations (\ref{two}) and (\ref{three}) 
are satisfied simultaneously. The optimization procedure in the spherical geometry has been implemented by \cite{tilaye09}, for restricted 
area with large field of views. Later \cite{Tilaye:2011} modified the objective functional of the optimization method \citep{Wiegelmann:2010}
for spherical geometry as
\begin{equation}L=L_{f}+L_{d}+\nu L_\mathrm{photo} \label{4}
\end{equation}
\begin{displaymath} L_\mathrm{f}=\int_{V}B^{-2}\big|(\nabla\times {\textbf{B}})\times
{\textbf{B}}\big|^2  r^2\sin\theta dr d\theta d\phi
\end{displaymath}
\begin{displaymath}L_\mathrm{d}=\int_{V}\big|\nabla\cdot {\textbf{B}}\big|^2
  r^2\sin\theta dr d\theta d\phi
\end{displaymath}
\begin{displaymath}L_\mathrm{photo}=\int_{S}\big(\textbf{B}-\textbf{B}_\mathrm{obs}\big)\cdot\textbf{W}(\theta,\phi)\cdot\big(
\textbf{B}-\textbf{B}_\mathrm{obs}\big) r^{2}\sin\theta d\theta d\phi
\end{displaymath}
where $L_{f}$ and $L_{d}$ measure how well the force-free Eqs.~(\ref{two}) and divergence-free (\ref{three}) conditions are fulfilled, 
respectively. The main reason for modification of the code was that we need to deal with boundary data of different noise levels and
qualities or even miss some data points completely. Hence, the third integral, $L_\mathrm{photo}$, is the surface integral over the 
photosphere which allows us to relax the field on the photosphere towards force-free solution without too much deviation from the original 
surface field data. 
%===============================================================================================================================================
\begin{figure*}[htp!]
   \centering
\subfigure[]{\includegraphics[viewport=0 0 588 575,clip,height=10.4cm,width=12.3cm]{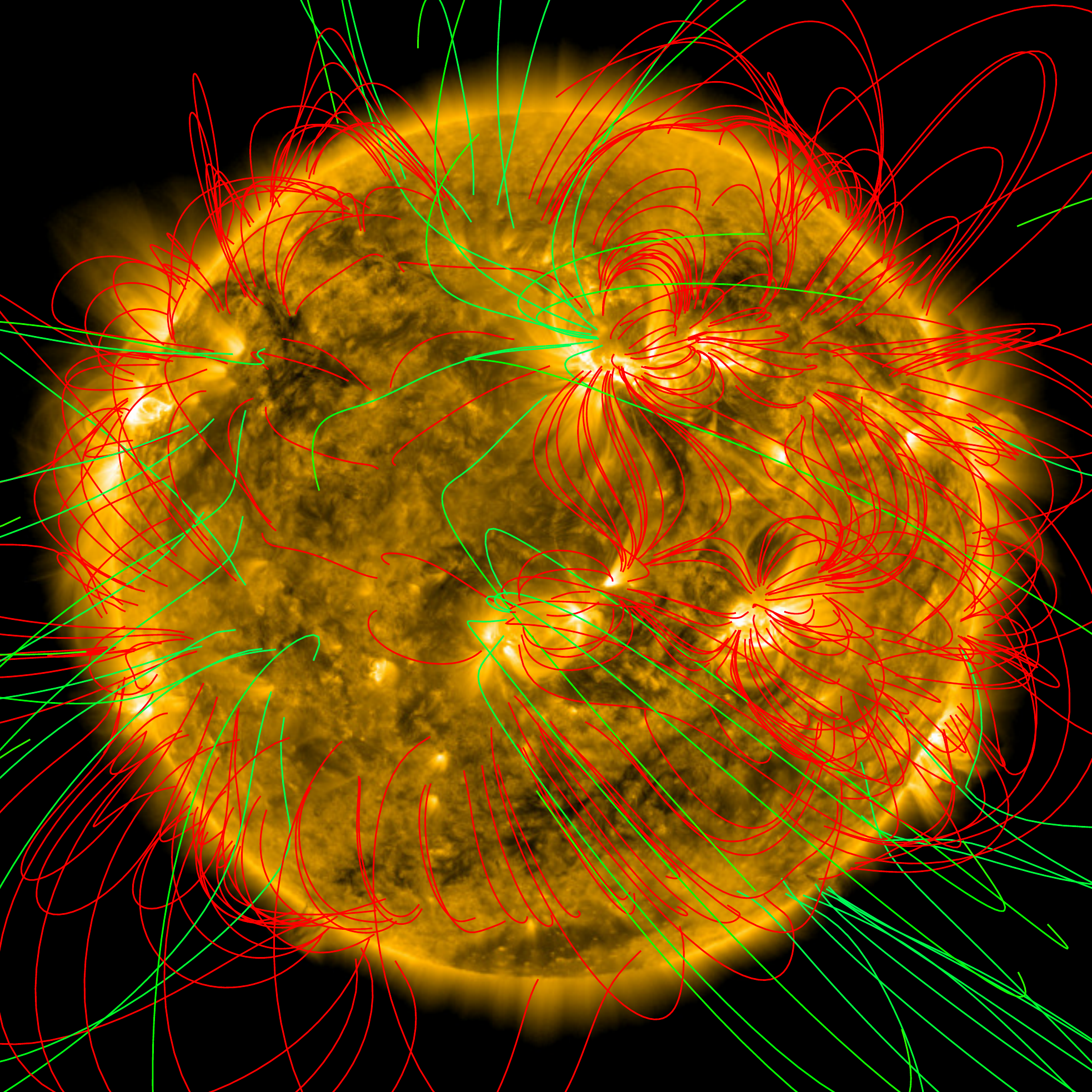}}
\subfigure[]{\includegraphics[viewport=0 0 588 575,clip,height=10.4cm,width=12.3cm]{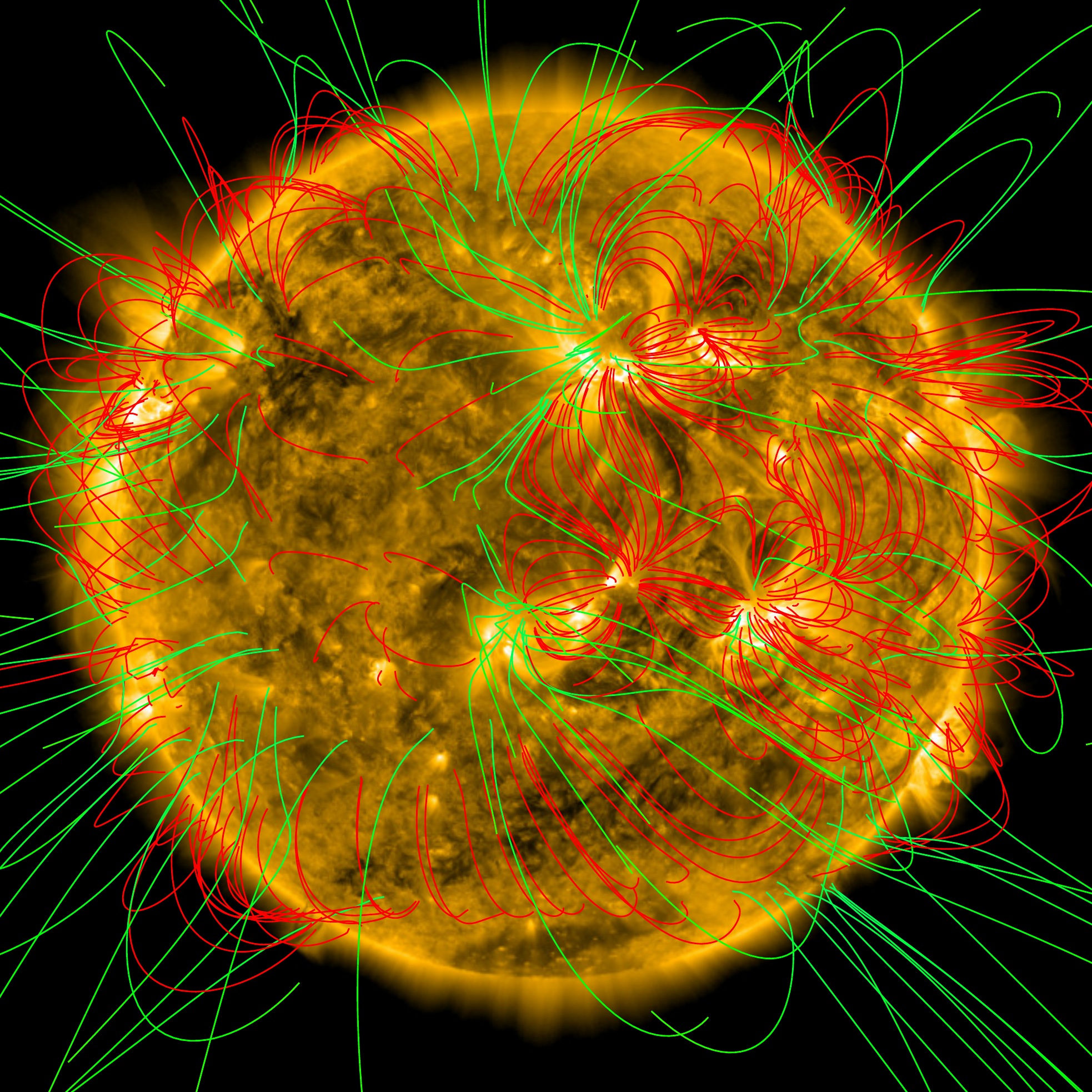}}
\caption{Global field lines of (a) the potential field model and (b) the NLFFF model overlaid on the AIA 171\AA{} image. 
Green and red lines represent open and closed magnetic field lines, respectively. }
\label{fig2}
 \end{figure*}
%===============================================================================================================================================

SOLIS/VSM provides full disk vector magnetograms from which Synoptic Carrington maps of the vector magnetic field components 
are synthesized. However, for pixels below the polarization threshold the inversion was not performed and field data there will 
be missing for these pixels (see Figure ~\ref{fig1}). Typically, the field is missing where its magnitude is small; thus these 
pixels would have a small impact on the model even if they were measured correctly. Within the error margin of a measured field 
value, any value is just as good as any other, and from this range of values we take the value that fits the force-free field best. 
In order to treat those pixels with missing data, we used the diagonal matrix, $\textbf{W}(\theta,\phi)$, 
which gives different weights to the observed surface field components depending on the relative accuracy in measurement. In this sense, 
missing data is considered most inaccurate and is taken into account by setting $\textbf{W}(\theta,\phi)$ to zero in all elements of
the matrix.

Photospheric magnetic field has a plasma-$\beta$ of order unity, which does not satisfy the force-free condition \citep{Gary:2001}. 
Therefore, the vector magnetogram data are inconsistent with the force-free assumption, which is absolutely essential condition for 
NLFFF extrapolation. To find suitable boundary conditions for the NLFFF field modeling, we have to preprocess the measured synoptic 
vector magnetograms by using a preprocessing scheme developed by \cite{tilaye09} in spherical geometry. This preprocessing scheme removes 
forces and torques from the boundary and approximates the photospheric magnetic field to the low plasma-$\beta$ force-free chromosphere. 
For a detailed description of the current code implementation, we refer to \cite{Wiegelmann07} and \cite{Tilaye:2011}.

%================================================================================================================================
\section{Results}
%================================================================================================================================
This study requires extrapolating the three-dimensional potential and NLFFF coronal fields from the photospheric boundary in global scale. 
We use synoptic maps of photospheric vector magnetic field observed during 4-31 March, 2012. During this observation there were about 25 active 
regions all over the solar globe. In order to use our spherical optimization code for global corona, we adopt a uniform spherical 
grid $r$, $\theta$, $\phi$ with $n_{r}=300$, $n_{\theta}=450$, and $n_{\phi}=900$ grid points in the direction of radius, latitude, 
and longitude, respectively, with the field of view of $[r_{\rm{min}}=1R_{\odot}:r_{\rm{max}}=2.5R_{\odot}]\times[\theta_{\rm{min}}=
-85^{\circ}:\theta_{\rm{max}}=85^{\circ}]\times[\phi_{\rm{min}}=0^{\circ}:\phi_{\rm{max}}=360^{\circ}]$. The code solves the NLFFF 
equations in the bounded domain between $1R_{\odot}$ and the source surface at $2.5R_{\odot}$. The outer boundary is kept fixed using 
the initial potential field values. All current-carrying field lines have to close inside the volume. The domain outside $2.5R_{\odot}$ is 
not included in the model, because the force-free approach is no longer justified here. The magnetic field extrapolation has been 
carried out almost for global corona by excluding the polar regions where the magnetic field measurements are highly being 
influenced by noise. Before performing NLFFF extrapolations, we use the preprocessed radial magnetic field component ${\bf {B}}_{r}$ 
to compute the corresponding potential field using a spherical harmonic expansion for initializing our spherical NLFFF code during 
relaxation towards a force-free state in the computational volume.
%===============================================================================================================================================
\begin{figure*}
\begin{center}
\includegraphics[viewport=0 162 850 420,clip,height=6.8cm,width=19.0cm]{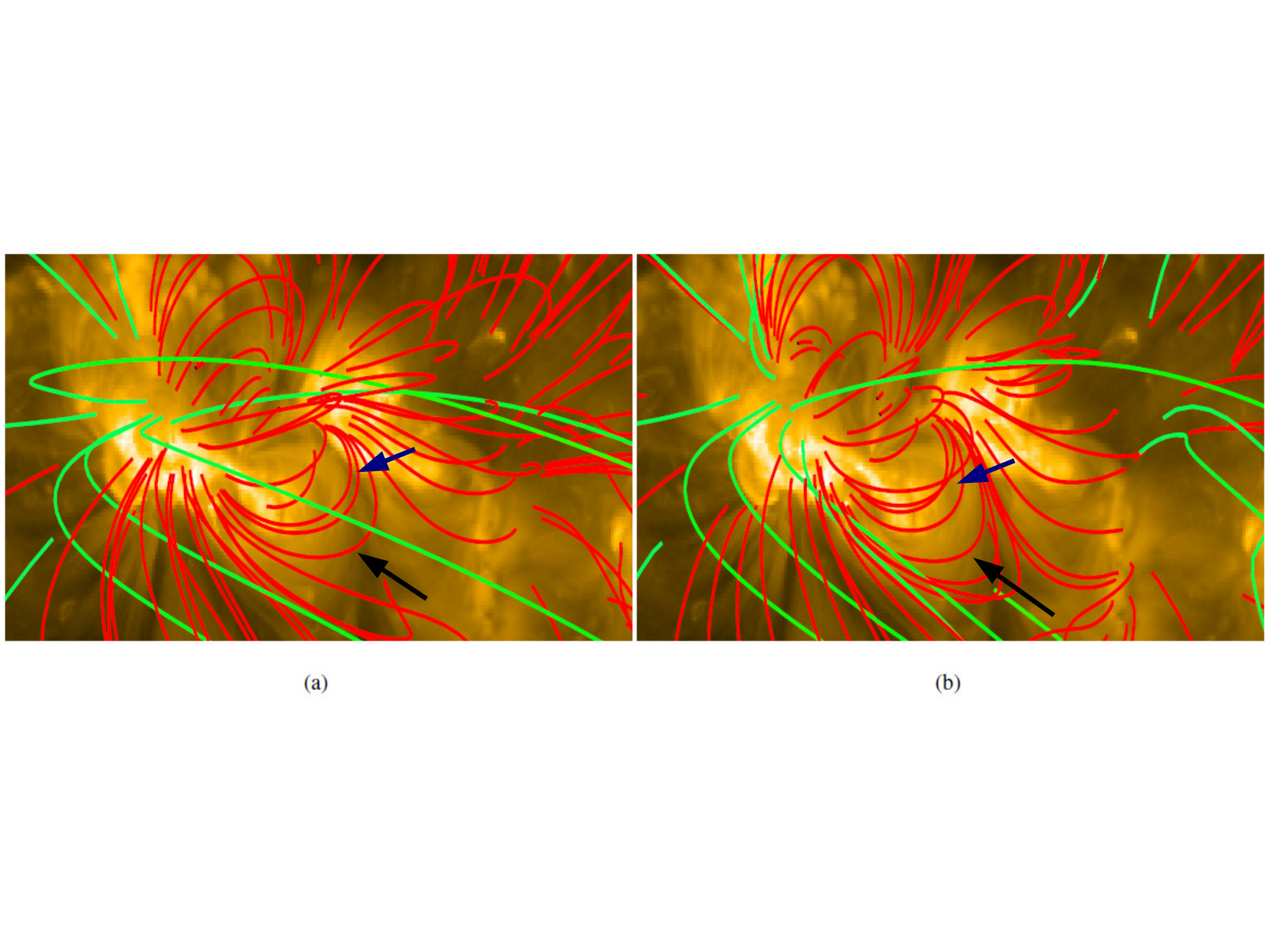}
\end{center}
\caption{Field lines of (a) the potential field model and (b) the NLFFF model around ARs 11429 and 11430 overlaid on the AIA 171\AA{} image. 
Green and red lines represent open and closed magnetic field lines, respectively.}
\label{fig3}
\end{figure*}
%===============================================================================================================================================
\begin{figure*}
\begin{center}
\subfigure[]{\includegraphics[viewport=50 50 558 555,clip,height=6.4cm,width=7.3cm]{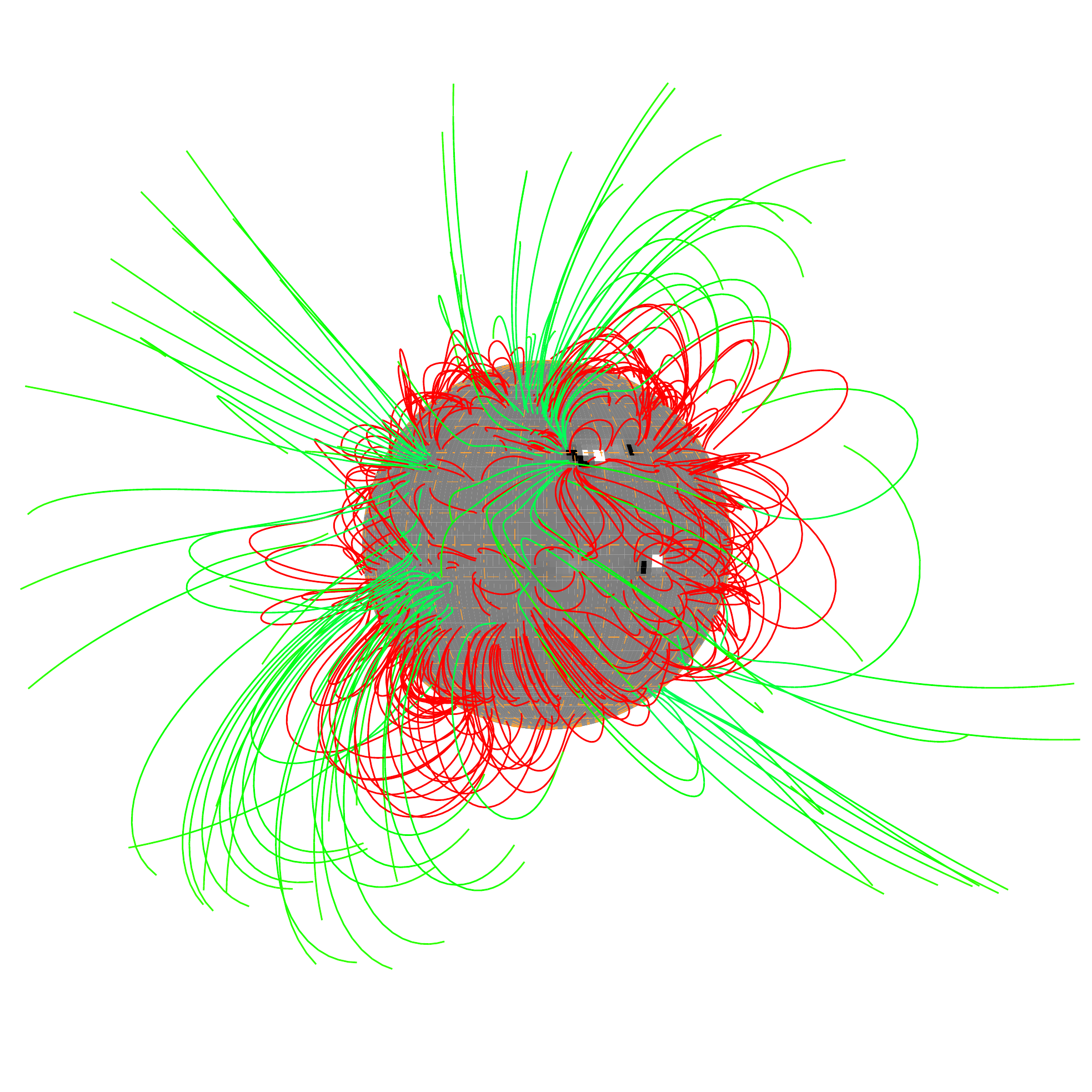}}
\subfigure[]{\includegraphics[viewport=20 20 550 550,clip,height=6.4cm,width=7.3cm]{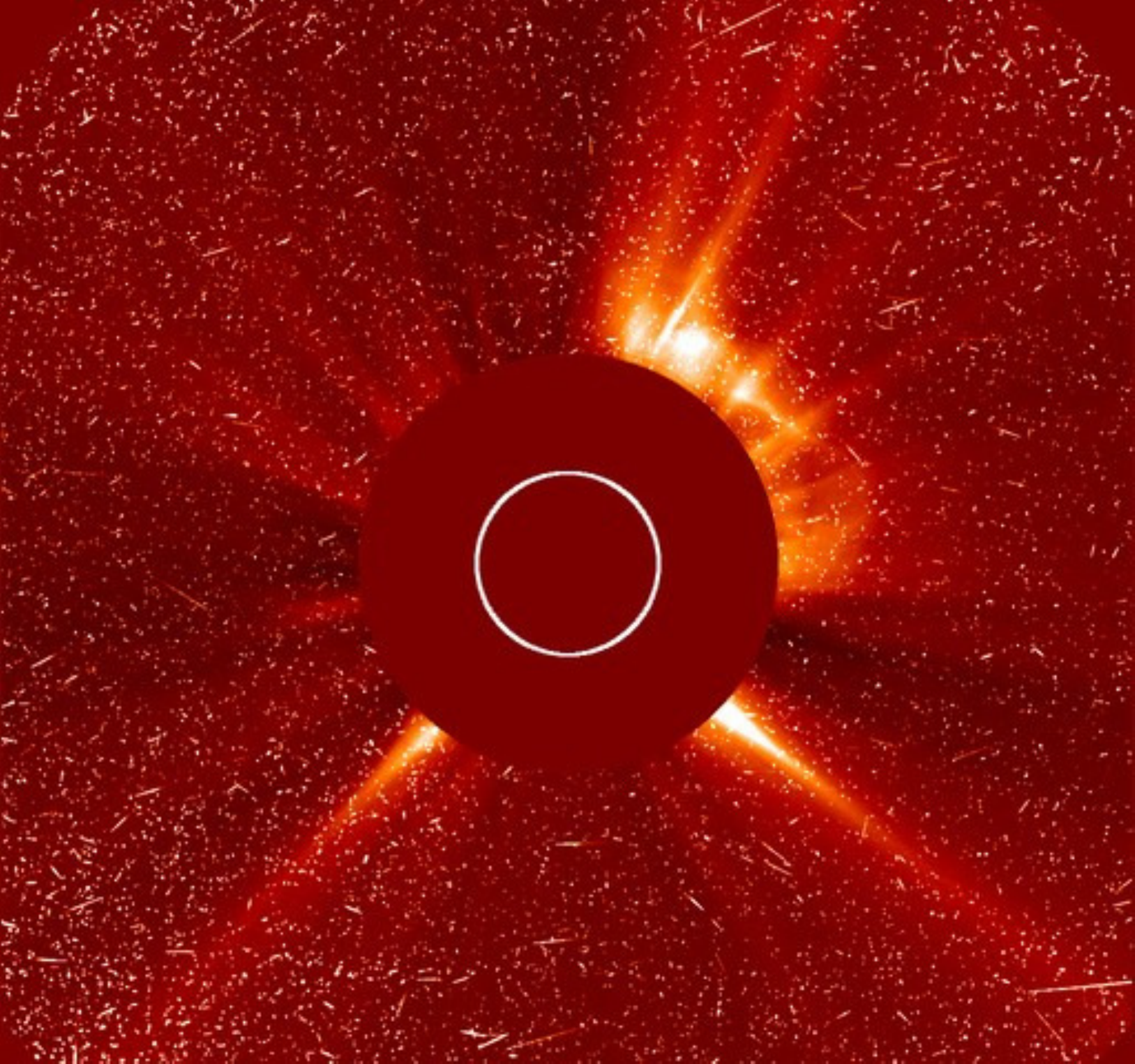}}
\end{center}
\caption{The magnetic field line skeletons (a) of the entire solar globe from the NLFFF model and image of the sun observed by 
SOHO/LASCO C2 coronagraph at 16:33UT.}
\label{fig3b}
\end{figure*}
%==============================================================================================================================================

The main purpose of this work is to study the structures of the global potential and NLFFF magnetic fields and to estimate free magnetic 
energy available to power solar eruptions during Carrington rotation 2121. In addition, we compare which of those two models is best agree  
with observation in global environment. To do this, we plot the selected fieldlines of the potential and NLFFF models in Figure~\ref{fig2}. 
We overlay the field lines with an AIA 171 \AA{} image. The field lines of the potential and NLFFF models are reconstructed from the same 
footpoints. The potential field lines in Figure~\ref{fig3}(a) have an obvious deviation from the observed EUV loops, since the projection 
of the field lines indicated by blue and black arrows divate from EUV loops. However, those loops are best overlaid by NLFFF lines than 
potential ones (see Figure~\ref{fig3}(b)). Therefore, the qualitative comparison between the model magnetic field lines and the observed 
EUV loops indicates that the NLFFF model provides a more consistent field for global corona magnetic field reconstruction. Figure~\ref{fig3b} 
shows that there is over all similarity between the corresponding NLFFF model field lines and image of the sun observed by SOHO/LASCO C2 coronagraph. 
However, one can see that NLFFF does not represent well a linear structure above coronal helmet in low-left corner in Figure~\ref{fig3}(b). 
This could be due to the missing data from the polar region. 
%===============================================================================================================================================
\begin{figure}
\begin{center}
\includegraphics[viewport=25 35 785 590,clip,height=6.2cm,width=9.8cm]{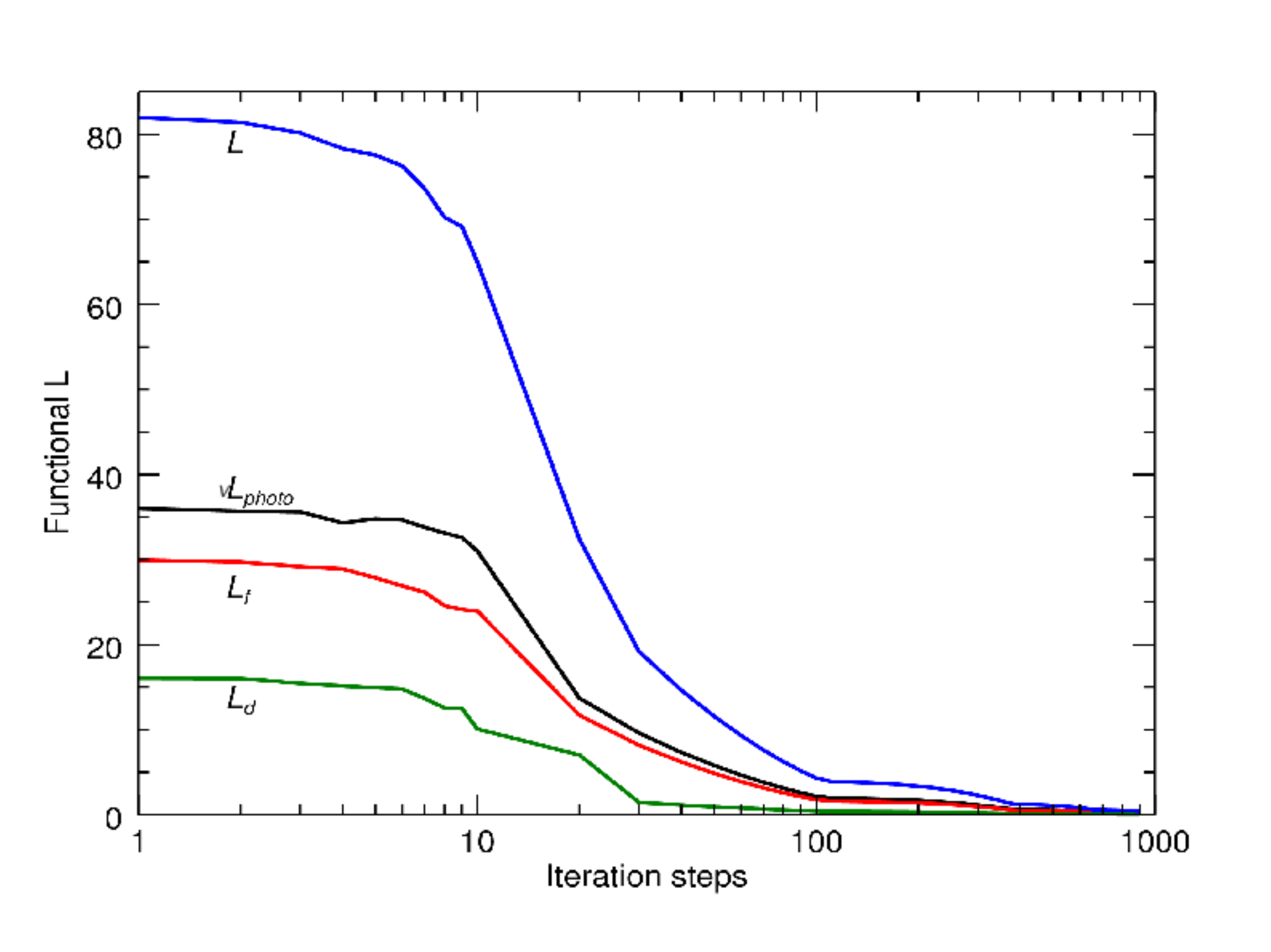}
\end{center}
\caption{Evolution of the entire functional L (blue line) and its three term in Eq.~\ref{4} during the optimization process. 
The black line corresponds to $L_{\rm photo}$, the red line to $L_{\rm f}$, and the green line to $L_{\rm d}$.}
\label{fig4}
\end{figure}
%===============================================================================================================================================

In addition to the above comparison to quantify the degree of disagreement between the two model vector field solutions in the 
global corona volume that are specified on the identical sets of grid points, we use the vector correlation metric ($C_{\rm vec}$) 
which is also used analogous to the standard correlation coefficient for scalar functions. The correlation was calculated \citep{Schrijver06} 
from
\begin{equation}
C_\mathrm{ vec}= \frac{ \sum_i \textbf{v}_{i} \cdot \textbf{u}_{i}}{ \Big( \sum_i |\textbf{v}_{i}|^2 \sum_i
|\textbf{u}_{i}|^2 \Big)^{1/2}}, \label{6}
\end{equation}
where $\textbf{v}_{i}$ and $\textbf{u}_{i}$ are the vectors at each grid point $i$. If the vector fields are identical, then $C_{\rm vec}=1$; 
if $\textbf{v}_{i}\perp \textbf{u}_{i}$ , then $C_{\rm vec}=0$. The degree of convergence towards a force-free and divergence-free model 
solution can be quantified by the integral measures of the Lorentz force and divergence terms in the minimization functional in Equation (\ref{4}), 
computed over the entire solar globe. $L_{\rm f}$ and $L_{\rm d}$ of Equation (\ref{4}) measure how well the force-free and divergence-free 
conditions are fulfilled, respectively. In Table~\ref{table1}, we provide some quantitative measures to rate the quality of our reconstruction. 
Column $1$ names the corresponding models. Columns $2-3$ show how well the force-balance and solenoidal conditions are fulfilled for both models. 
Figure ~\ref{fig4} shows how well the functional $L$ converge to zero during iteration process. In the last column, the vector correlation 
shows that there is disagreement between the two model field solutions.  

\begin{figure}
\begin{center}
\includegraphics[viewport=175 65 665 320,clip,height=4.3cm,width=9.2cm]{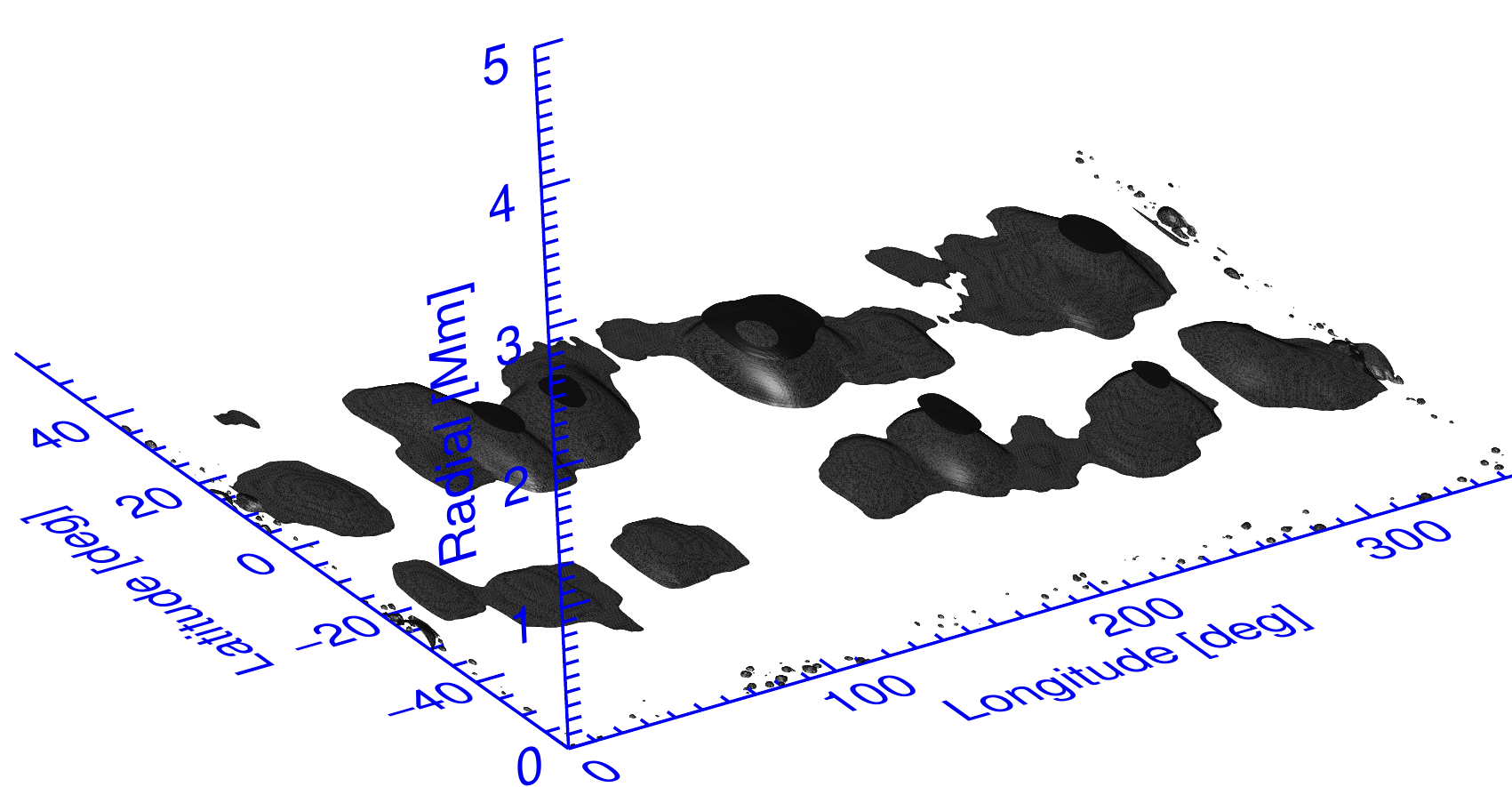}
\end{center}
\caption{Iso-surfaces (ISs) of free magnetic energy density ($8.5\times10^{20}$ erg) computed within the entire computational domain.}
\label{fig6}
\end{figure}

The energy stored in the magnetic field as a result of field line stressing into a nonpotential configuration has been identified as the 
source of flare energy. Therefore, to understand the physics of solar flares, including the local reorganisation of the magnetic field 
and the acceleration of energetic particles, we have first to estimate the free magnetic energy available for such phenomena. This free 
magnetic energy can be converted into kinetic and thermal energy. We estimate the free magnetic energy, the difference between the extrapolated NLFFF and 
the potential field with the same normal boundary conditions in the photosphere. We therefore estimate the upper 
limit to the free magnetic energy associated with coronal currents of the form
\begin{equation}
   E_\mathrm{free}=\frac{1}{8\pi}\int_{V} \Big({B^2}_{\mbox{nlff}}-{B^2}_{\mbox{pot}}\Big)dV\,\,  \label{a}
\end{equation}
${B}_{\mbox{pot}}$ and ${B}_{\mbox{nlff}}$ represent the potential and NLFFF magnetic field, respectively.
Our result for the estimation of free-magnetic energy in Table~\ref{table2} shows that the NLFFF model has $10.3\%$ more energy 
than the corresponding potential field model. Figure~\ref{fig6} shows iso-surface plot of free magnetic energy density in the volume 
above synoptic map. There are strong free energy concentrations above each active region over the solar globe.

%===============================================================================================================================================
%===============================================================================================================================================
\begin{table}
\caption{Evaluation of the reconstruction quality for the potential field and NLFFF models. We have used spherical grids of $300 \times 450 \times 900$.}
\label{table1}
\centering
\begin{tabular}{ccccc}   
\hline \hline Model &$L_{\rm f}$&$L_{\rm d}$ &$L_{\rm photo}$ &$C_{\rm vec}$ \\
\hline
%&\multicolumn{3}{c}{Spherical grid $225 \times 375 \times 425$} &&&& \\
Potential &$0.000$& $ 0.000$& $ 0.001$&$ 1$ \\
 NLFFF &$0.391$&$0.697$& $ 0.302$&$0.893$ \\
\hline
\end{tabular}
\end{table}
%==============================================================================================================================================
%==============================================================================================================================================
\begin{table}
\caption{The magnetic energy associated with the 3-D potential and NLFFF field configurations calculated from synoptic vector magnetogram.}
\label{table2}
\centering
\begin{tabular}{ccc}
 \hline \hline
Model & $E_{\rm{total}}(10^{33}{\rm{erg}})$& $E_{\rm{free}}(10^{33}{\rm{erg}})$\\
\hline
Potential &$41.6$&$0$\\
NLFFF &$46.4$&$4.8$\\
\hline
\end{tabular}
\end{table}

%==============================================================================================================================================
%================================================================================================================================
\section{Conclusion and outlook}
%================================================================================================================================
 Most of the NLFFF procedures are implemented in the Cartesian coordinates. Therefore, both potential and nonlinear force-free 
 field (NLFFF) codes in Cartesian geometry are not well suited for larger domains, since the spherical nature of the solar surface 
 cannot be neglected when the field of view is large. Therefore, it is necessary to implement a NLFFF procedure in spherical geometry 
 for use when large-scale boundary data are in use. 

In this study, we have investigated the coronal magnetic field and free magnetic energy associated with global corona by analyzing Carrington 
synoptic maps of photospheric vector magnetic field synthesized from Vector Spectromagnetograph (VSM) on Synoptic Optical Long-term 
Investigations of the Sun (SOLIS) which has been observed during 4-31 March, 2012. The Carrington rotation number for this observation is 2121.
During this particular observation, there were about ten active regions distributed across the globe. We have used our spherical NLFFF 
and potential codes to compute the magnetic field solutions over global corona. This is the first NLFFF magnetic field extrapolation for 
the global corona using real data.

We have compared the magnetic field solutions from both potential and NLFFF models. The qualitative comparison between the model magnetic 
field lines and the observed EUV loops indicates that the NLFFF model provides a more consistent field for global corona magnetic field 
reconstruction. For this particular Carrington rotation we find that the global NLFFF magnetic energy density is $10.3\%$ higher than the 
potential one. For future, we have a plan to study the evolution of global free magnetic energy. In this study, most of this free energy 
is located in active regions. 

%================================================================================================================================
\begin{acknowledgements} This work utilizes SOLIS data obtained by the NSO Integrated Synoptic Program (NISP), managed by the 
National Solar Observatory, which is operated by the Association of Universities for Research in Astronomy(AURA), Inc. under a 
cooperative agreement with the National Science Foundation. This research was supported by an appointment to the NASA Postdoctoral 
Program at the Goddard Space Flight Center (GSFC), administered by Oak Ridge Associated Universities through a contract with NASA. 
The work of T. Wiegelmann was supported by DLR-grant $50$ OC $453$  $0501$.
\end{acknowledgements}
%================================================================================================================================

%================================================================================================================================


\begin{thebibliography}{}
\bibitem[Amari et~al.(2013)]{Amari:2013} Amari, T., Aly, J.~J., Canou, A. et al., 2013, \aap, {\bf 553}, A43
\bibitem[Aschwanden(2012)]{Aschwanden:2012} Aschwanden, M.~J., 2012, \solphys, preprint doi:10.1007/s11207-012-0203-6
\bibitem[Balasubramaniam et al.(2011)]{Balasubramaniam0:2011} Balasubramaniam, K.~S., Pevtsov, A.~A. \& Cliver, E.~W., et al., 2011, \apj, {\bf 743}, 202 
\bibitem[Balasubramaniam \& Pevtsov(2011)]{Balasubramaniam:2011} Balasubramaniam, K.~S. \& Pevtsov, A., 2011, 
in Society of Photo-Optical Instrumentation Engineers (SPIE) Conference Series, Vol. {\bf 8148}
\bibitem[DeRosa et al.(2009)]{DeRosa:2009} DeRosa, M.~L., Schrijver, C. J., Barnes, G., et al., 2009, \apj, {\bf 696}, 1780 
\bibitem[Fang et al.(2012)]{Fang:2012} Fang, F., Manchester, IV, W., Abbett, W.~P., et al., 2012, \apj, {\bf 754}, 15
\bibitem[Forbes(2000)]{Forbes:2000} Forbes, T.~G., 2000, \jgr, {\bf 105}, 23153	
\bibitem[Gary(2001)]{Gary:2001} Gary, G.~A., 2001, \solphys, {\bf 203}, 71 
\bibitem[Gosain et al.(2013)]{Gosain:2013} Gosain, S., Pevtsov, A.~A., Rudenko, G.~V., et al., 2013, \apj, {\bf 772}, 52
\bibitem[Guo et al.(2008)]{Guo:2008} Guo, Y., Ding, M. D., Wiegelmann, T., et al., 2008, \apj, {\bf 679}, 1629
\bibitem[Guo et al.(2012)]{Guo:2012} Guo, Y., Ding, M. D., Liu, Y., et al., 2012, \apj, {\bf 760}, 47
\bibitem[Henney et al.(2006)]{Henney:2006} Henney, C.~J., Keller, C.~U. \& Harvey, J.~W., 2006, 
in Astronomical Society of the Pacific Conference Series, Vol. {\bf 358}, 92
\bibitem[Jing et al.(2010)]{Jing:2010} Jing, J., Tan, C., Yuan, Y., et al., 2010, \apjl, {\bf 713}, 440
\bibitem[Jones et al.(2002)]{Jones:2002} Jones, H.~P, Harvey, J.~W., Henney, C.~J., et al., 2002,in ESA Special Publication, SOLMAG 2002. Proceedings
of the Magnetic Coupling of the Solar Atmosphere Euroconference,Vol. {\bf 505}, 15
\bibitem[Keller et al.(2003)]{Keller:2003} Keller, C.~U., Harvey, J.~W. \& Giampapa, M.~S., 2003, 
in Society of Photo-Optical Instrumentation Engineers (SPIE) Conference Series, Vol. {\bf 4853}, 194
\bibitem[Lin et al.(2000)]{Lin:2000} Lin, H., Penn, M.~J. \& Tomczyk, S., 2000, \apjl, {\bf 541}, L83 
\bibitem[Lin et al.(2004)]{Lin:2004} Lin, H., Kuhn, J.~R. \& Coulter, R., 2004, \apjl, {\bf 613}, L177 
\bibitem[Liu \& Lin(2000)]{Liu:2008} Liu, Y. \& Lin, H., 2008, \apj, {\bf 680}, 1496
\bibitem[Low \& Lou(1990)]{Low:1990} Low, B.~C. \& Lou, Y.~Q., 1990, \apj, {\bf 352}, 343
\bibitem[Low(2001)]{Low:2001} Low, B.~C., 2001, \jgr, {\bf 106}, 25141
\bibitem[Malanushenko et~al.(2012)]{Malanushenko:2012} Malanushenko, A., Schrijver, C. J., DeRosa, M. L., et al., 2012, \apj, {\bf 756}, 153
\bibitem[Martens et~al.(2012)]{Martens:2012} Martens, P.~C., Attrill, G. D. R., Davey, A. R., et al., 2012, \solphys, {\bf 275}, 79
\bibitem[Metcalf et al.(2005)]{Metcalf:2005} Metcalf, T.~R., Leka, K.~D. \& Mickey, D.~L., 2005, \apjl, {\bf 623}, L53  
\bibitem[Meyer et~al.(2013)]{Meyer:2013} Meyer, K.~A., Sabol, J., Mackay, D. H., et. al., 2013, \apjl, {\bf 770}, L18
\bibitem[Pevtsov(2000)]{Pevtsov:2000} Pevtsov, A.~A., 2000, \apj, {\bf 531}, 553
\bibitem[Pevtsov \& Acton(2001)]{Pevtsov:2001} Pevtsov, A.~A., \& Acton, L.~W., 2001, \apj, {\bf 554}, 416
\bibitem[R{\'e}gnier \& Canfield(2006)]{Regnier:2006} R{\'e}gnier, S. \& Canfield, R.~C., 2006, \aap, {\bf 451}, 319
\bibitem[R{\'e}gnier(2007)]{Regnier:2007} R{\'e}gnier, S., 2007, \apjl, {\bf 669}, L53
\bibitem[Rudenko \& Anfinogentov(2011)]{Rudenko:2011} Rudenko, G.~V. \& Anfinogentov, S.~A., 2011, arXiv:1104.1228
\bibitem[Schatten et al.(1969)]{Schatten:1969} Schatten, K.~H., Wilcox, J.~M. \& Ness, N.~F., 1969, \solphys, {\bf 6}, 442
\bibitem[Schrijver and DeRosa.(2003)]{Schrijver:2003} Schrijver, C.~J. \& DeRosa, M.~L., 2003, \solphys, {\bf 212}, 165 
\bibitem[Schrijver et al.(2006)]{Schrijver06} Schrijver, C.~J., Derosa M.~L., Metcalf T. R., et. al., 2006, \solphys, {\bf 235}, 161
\bibitem[Schrijver et al.(2008)]{Schrijver:2008} Schrijver, C.~J., et. al., 2008, \apj, {\bf 675}, 1637 
\bibitem[Skumanich \& Lites(1987)]{Skumanich:1987} Skumanich, A. \& Lites, B.~W., 1987, \apj, {\bf 322}, 473
\bibitem[Tadesse et~al.(2009)]{tilaye09} Tadesse, T., Wiegelmann, T. \& Inhester, B., 2009, \aap, {\bf 508}, 421
\bibitem[Tadesse et~al.(2011)]{Tilaye:2011} Tadesse, T., Wiegelmann, T., Inhester, B., et al., 2011, \aap, {\bf 527}, 30
\bibitem[Tadesse et~al.(2012a)]{Tilaye:2012} Tadesse, T., Wiegelmann, T., Inhester, B., et al., 2012, \solphys, {\bf 281}, 53
\bibitem[Tadesse et~al.(2012b)]{Tilaye:2012a} Tadesse, T., Wiegelmann, T., Inhester, B., et al., 2012, \solphys, {\bf 277}, 119
\bibitem[Tadesse et~al.(2013a)]{Tadesse:2013a} Tadesse, T., Wiegelmann, T., Inhester, B., et al., 2013, \aap, {\bf 550}, 14
\bibitem[Tadesse et~al.(2013b)]{Tilaye:2013b} Tadesse, T., Wiegelmann, T., MacNeice, P.~J., et al., 2013, \solphys, {\bf 000}, 00
\bibitem[Welsch(2006)]{Welsch2006} Welsch, B.~T., 2006, \apj, {\bf 638}, 1101
\bibitem[Wheatland et~al.(2000)]{Wheatland:2000} Wheatland, M.S., Sturrock, P.A., \& Roumeliotis, G., 2000, \apj, {\bf 540}, 1150
\bibitem[Wheatland \& R{\'e}gnier(2009)]{Wheatland:2009} Wheatland, M.S. \& R{\'e}gnier, S., 2009, \apjl, {\bf 700}, 88
\bibitem[Wheatland \& Leka(2011)]{Wheatland:2011} Wheatland, M.S. \& Leka, K.D., 2011, \apj, {\bf 728}, 112
\bibitem[Wiegelmann(2004)]{Wiegelmann04} Wiegelmann, T., 2004, \solphys, {\bf 219}, 87
\bibitem[Wiegelmann(2007)]{Wiegelmann07} Wiegelmann, T., 2007, \solphys, {\bf 240}, 227
\bibitem[Wiegelmann \& Inhester(2010)]{Wiegelmann:2010} Wiegelmann, T. \& Inhester, B., 2010, \aap, {\bf 516}, A107
\bibitem[Wiegelmann \& Sakurai(2012)]{Wiegelmann:2012W} Wiegelmann, T. \& Sakurai, T., 2012, Living Reviews in Sol. Phys., {\bf 9}, 5
\bibitem[Wiegelmann et al.(2012)]{Wiegelmann:2012} Wiegelmann, T., et al., 2012, \solphys, {\bf 281}, 37
\end{thebibliography}
\end{document}